%% file: prldraft.tex
\documentclass[aps,prd,twocolumn,showpacs,superscriptaddress,groupedaddress,amsmath,nofootinbib]{revtex4-1}  
\usepackage{graphicx}  
\usepackage{dcolumn}   
\usepackage{bm}        
\usepackage{amssymb}
\usepackage{url}
\usepackage{subfigure}
\usepackage{hyperref}
\usepackage{cancel}
\usepackage{natbib}
\usepackage{ulem}
\usepackage{aas_macros}
\def \fermi{{\it Fermi}-LAT~}

\begin{document}


\title{Probing the \textquotedblleft Sea" of Galactic Cosmic Rays with \fermi} 
\author{Felix Aharonian}\email{felix.aharonian@mpi-hd.mpg.de}\affiliation{Dublin Institute for Advanced Studies, 31 Fitzwilliam Place, Dublin 2, Ireland\\Max-Planck-Institut f{\"u}r Kernphysik, P.O. Box 103980, 69029 Heidelberg, Germany} 
\author{Giada Peron}\email{peron@mpi-hd.mpg.de}\affiliation{Max-Planck-Institut f{\"u}r Kernphysik, P.O. Box 103980, 69029 Heidelberg, Germany}
\author{Ruizhi Yang}\email{ryang@mpi-hd.mpg.de}
\affiliation{Max-Planck-Institut f{\"u}r Kernphysik, P.O. Box 103980, 69029 Heidelberg, Germany}
\author{Sabrina Casanova}\email{sabrinacasanova@gmail.com}
\affiliation{Max-Planck-Institut f{\"u}r Kernphysik, P.O. Box 103980, 69029 Heidelberg, Germany \\ Institute of Nuclear Physics, Radzikowskiego 152, 31-342 Krakow, Poland
}
\author{Roberta Zanin}\email{roberta.zanin@mpi-hd.mpg.de}
\affiliation{Max-Planck-Institut f{\"u}r Kernphysik, P.O. Box 103980, 69029 Heidelberg, Germany}

\date{Received:  / Accepted: } 

\begin{abstract}

High energy $\gamma$ rays from Giant Molecular Clouds (GMCs) carry direct information about the spatial and energy distributions of Galactic Cosmic Rays (CRs). The recently released catalogs of GMCs contain sufficiently massive clouds to be used as barometers for probing, through their $\gamma$-ray emission,  the density of CRs throughout the Galactic Disk.  Based on the data of \fermi{}, we report the discovery of $\gamma$-ray signals from nineteen GMCs located at distances up to 12.5 kpc.   The galactocentric radial distribution of the CR density derived from the $\gamma$-ray and CO observations  of these objects, as well as from some nearby clouds that belong to the Gould  Belt complex,  unveil a homogeneous \textquotedblleft sea" of CRs with a constant density and spectral shape close to the flux of directly (locally) measured CRs.   {  We found noticeable deviations from the \textquotedblleft sea level"  only in some locations characterized by enhanced  CR density in the galactocentric 4--6 kpc ring. Furthermore, we found a hint for fluctuations of the CR density in different locations within the same 4--6 kpc ring.  The confirmation of this result with the next-generation $\gamma$-ray detectors based on the higher quality data and denser coverage of galactocentric distances, would have dramatic implications for the understanding of the origin of Galactic CRs.}

\date{}

\end{abstract}
\maketitle
%




\maketitle



\input{main}
\bibliography{Bibliography}
\appendix{}
\include{Appendix}

\end{document}

%% file: main.tex
\section{Introduction}

The recent years have been marked by impressive progress in the precision and quality of direct Cosmic Ray (CR) measurements.  Yet, the key issues concerning the origin of Galactic CRs are not fully understood and resolved.

A breakthrough in the field is expected from  $\gamma$-ray observations. This concerns both the acceleration and propagation aspects of CR studies. While the detection and identification of $\gamma$-ray sources unveil the sites of CR production, the diffuse $\gamma$-ray emission of the Galactic Disk contains  information about the spatial and energy distributions of CRs in the Milky Way. The accumulation and effective mixture of relativistic particles through their convection and diffusion in the interstellar magnetic fields results in the formation of the so-called {\it sea} of Galactic CRs. The level and the energy spectrum of the CR sea is determined by the operation of all Galactic accelerators over the confinement time of CRs.  At low (GeV) energies it is estimated about  $10^7$ years, and decreases with energy as $E^{-\delta}$,  with $\delta \sim 0.3 -- 0.5$.   Since the effective \textquotedblleft lifetimes"  of potential CR factories are shorter than the CR confinement time,  one should expect a rather homogeneous distribution of CRs on large (kpc) scales. The homogeneity of distribution of CRs, however, can be violated on smaller scales,  in particular in the proximity of recent or currently operating powerful particle accelerators.

The diffuse Galactic $\gamma$ radiation is produced by relativistic electrons, protons and nuclei interacting with the interstellar gas and radiation fields.   The energy interval from 100 MeV to 100 GeV is dominated by $\gamma$-rays from the decays of secondary $\pi^0$ mesons  \citep{stecker77, aharonianatoyan, SMR2004}.  
{  Thus, the diffuse gamma-ray emission contains essential information about the  {\it mean}  density of Galactic CRs  averaged along the line of sight:  $\rho_{\rm CR} \propto F_\gamma /N_{\rm H}$, where  $F_\gamma$  and $N_{\rm H}$ are the $\gamma$-ray flux and the gas column density in the given direction of ISM, respectively.  This emission has been analyzed, for example, in refs. \cite{strong1996gradient,acero2016development,yang16}, where rings of gas at different distances from the Galactic Center have been considered. This method has certain limitations. First of all, the measured CR density is the mean value averaged over the vast areas of the rings (typically, $\gtrsim$ 10 kpc$^2$  ).  Therefore it can provide only integral information about the CR density.  Secondly, the method assumes cylindrical symmetry, which perhaps could be considered a reasonable approximation but yet remains a {\it ad hoc} assumption.  In fact it is not apparent that the relevant parameters like the CR diffusion coefficient, the spatial distribution of CR accelerators, and hence ultimately the CR density, do not vary on very large (multi-kpc) scales within the galactocentric rings. 
Moreover, the measurement of the CR density through the diffuse gamma-ray emission is affected by the contamination from the diffuse Inverse Compton emission, and unresolved $\gamma$-ray sources. Finally, 
the gas column density is dominated by the contribution of a limited number of Giant Molecular Clouds (GMCs), thus the derived CR density corresponds to the value averaged over specific locations occupied by these clouds.} 

{  In this context, $\gamma$-rays from {\it individual} GMCs do provide straightforward and localized (differential) information about CRs. Moreover, the high gas density of GMCs and their compactness make negligible the contribution from the inverse Compton component of diffuse background radiation and significantly reduces the level of potential contamination from other large-scale sources.} Thus, these GMCs can be treated as unique CR barometers distributed throughout the Galaxy \cite{aharonian2001gamma,casanova2010molecular}.  The case of  {\it passive} GMCs, i.e., the clouds located far from the active accelerators, is of particular interest. 
The detection of  $\gamma$ rays from such clouds tells us about the level of the CR sea without substantial contamination by particles injected by nearby objects. 

{  Under the assumption that CRs freely penetrate the cloud}, the flux of $\gamma$-rays from a \textquotedblleft passive" cloud depends  on a  single parameter, the ratio $M/d^2$,  where $M$ is the clouds mass and $d$ is the distance to the source, namely: 
\begin{equation}\label{eq:expflux}
{ F_\gamma(E_\gamma) =  \frac{M}{ d^2}  \frac{\xi_N}{m_p} \int \mathrm{d}E_p \frac{\mathrm{d}\sigma}{\mathrm{d}E_\gamma}F_p (E_p) }; 
\end{equation}
Here $m_p$ is the proton mass and the parameter $\xi_N$ takes into account the contribution of nuclei to the $\gamma$-ray flux; we considered $\xi_N \approx 1.8$  as calculated for the standard composition of the interstellar medium and CRs \cite{kafexhiu14}. $F_p(E_p)$ is the spectrum of CR protons. For comparison, we use the flux reported by the AMS collaboration \cite{ams_proton}. The latter is well described, as a function of rigidity, $R$ , above 45 GV ($\sim$45 GeV) by the equation:

\begin{equation} F_p(R)= C \bigg( \frac{R}{45 \ \mathrm{GV}} \bigg)^\gamma \bigg[1 + \bigg( \frac{R}{R_0} \bigg)^{\frac{\Delta \gamma}{s}} \bigg]^s  
\end{equation} with $\gamma= -2.849$, $R_0$=336 GV, $\Delta \gamma=0.133$ and $s=0.024$ that takes into account the recently discovered hardening at $\sim$200 GeV. For the differential cross-section of $pp$ interactions, $\frac{\mathrm{d}\sigma}{\mathrm{d}E_\gamma}$, we use the parametrization  from ref.\cite{kafexhiu14}. 

\begin{figure}[h!]
\includegraphics[width=\linewidth]{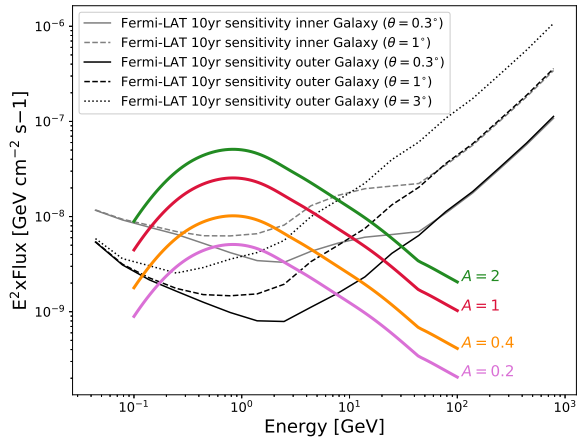}
\caption{Gamma-ray fluxes calculated for different values of the $A$ parameter, For comparison, the \fermi flux sensitivities for different regions of the sky and for different angular extensions of sources are shown.} 
\label{fig:fluxvssens1}
\end{figure}

It is convenient \cite{aharonian2001gamma} to write the ratio $M/d^2$  in the normalized form 
\begin{equation}
A =  M_5/d_{\rm kpc}^2 , 
\label{eq:A-parameter}
\end{equation}
where $M_5=M/10^5 M_\odot$  and $d_{\rm kpc}=d/1 \ \rm kpc$ and  use it to set the detection threshold of the \fermi{}.   
In Fig. \ref{fig:fluxvssens1}, we show the $\gamma$-ray fluxes calculated for  different values of $A$ against the 10-yr sensitivities of \fermi. The \textit{inner sensitivity} corresponds to the  minimum detectable flux calculated for $l,b$=(0$^\circ$,0$^\circ$)  and generally valid for the inner part of the Galactic Disk, namely  for $|l|\lesssim 60^\circ$ and $|b|\lesssim 5^\circ$. The  \textit{outer sensitivity}  corresponds to minimum detectable flux of sources located in the region $l,b$=(0$^\circ$,30$^\circ$) and characterizes actually the sensitivity for a significantly broader fraction of the Galactic Disk:  $|l|>60^\circ$, $5< |b|<45^\circ$). For details, see  the \fermi performances web page\footnote{\url{www.slac.stanford.edu/exp/glast/groups/canda/lat_Performance}}. The curves in Fig. \ref{fig:fluxvssens1} are calculated for different source angular extensions $\theta$ by multiplying the sensitivity for the point-like source by the factor $\sqrt[]{\sigma_{PSF}^2 +{\theta}^2}/\sigma_{PSF} $.  We deduce from Fig.\ref{fig:fluxvssens1} that in the case of sources with 
angular extensions smaller than  $1^{\circ}$,   \fermi is capable to detect molecular clouds with $A\gtrsim 0.4$. In the case of location of compact clouds in uncrowded regions,  the detection threshold can be as small as 0.2. On the other hand, for very close clouds,  $d \ll 1$~kpc,   $A$ should significantly exceed 1 to compensate the loss of the sensitivity of \fermi due to the large (several degrees)  extensions of clouds. In the opposite case of gas complexes located in the Galactic Center  (GC), the reduction of the flux ($\propto d^{-2}$)  is compensated by the vast masses, $M\sim10^7 M_\odot$, and small angular extensions of  $\sim 10 $ arcmin.  This explains why so far positive $\gamma$-ray signals have been reported only from the nearby Gould Belt clouds \cite{Neronov2017, yang2014probing, abdo2010fermi,ackermann2012fermiCLOUD} and from the  Sgr B complex in the GC  \cite{yang2015fermi}. The estimates of the CR density in these two essentially different parts of the Galaxy is very important but not sufficient for conclusions regarding  the overall distribution of CRs in the Milky Way.  

\begin{figure}[h!]
\includegraphics[width=\linewidth]{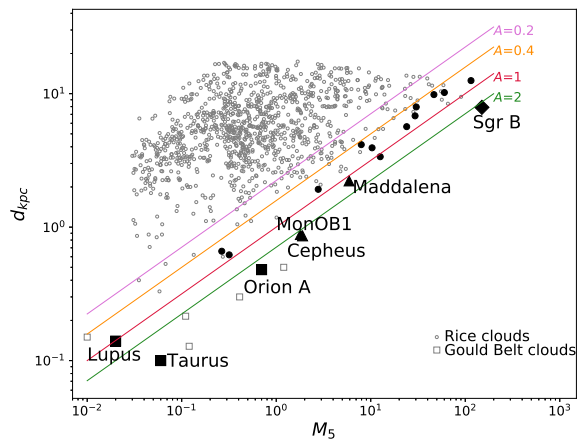}
\caption{Distribution of GMCs in the [$d_{kpc}$, $M_5$] plane. 
The detection thresholds corresponding to four different values of the A-parameter, are shown. The filled symbols  highlight the 19 clouds analyzed in this work. }
\label{fig:fluxvssens2}
\end{figure}

Clearly,  for probing the CR sea, a larger number of $\gamma$ ray emitting clouds, broadly distributed over the Galactic Plane, is needed.  The realization of this goal with \fermi could seem unrealistic, until recently, when a catalog of Galactic  GMCs, containing clouds of unexpectedly large masses has been released by Rice \cite{rice2016uniform}. In Fig. \ref{fig:fluxvssens2} we show the distribution of GMC of this catalog \cite{rice2016uniform} in the  [$d_{kpc}$, $M_5$] plane, together with the Gould Belt Clouds and some other close ($\lesssim$ 2 kpc) GMCs, namely Cepheus, Monoceros OB1 and Maddalena. We can see that the catalog of \cite{rice2016uniform} contains several clouds that have a value of $A$ high enough to be detected by \fermi, and are located at different distances from us .     

\begin{figure}[h!]
\includegraphics[width=\linewidth]{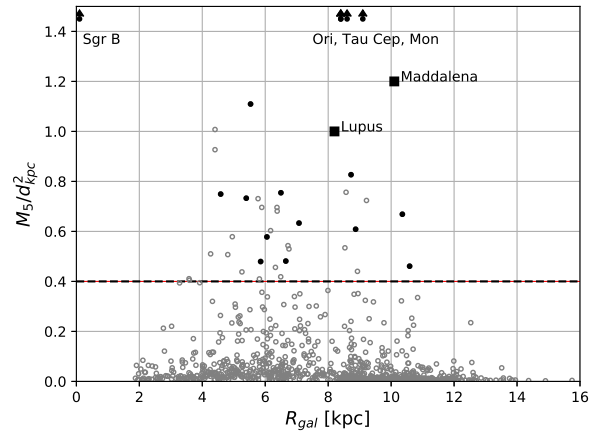}
\caption{Distribution of the $A$ ($M_5/d^2_{kpc}$) parameter of the GMCs of the considered catalog \cite{rice2016uniform}, as a function of the distance from the Galactic Center. The filled symbols indicate the clouds analyzed in this work.  Some of the nearby clouds, as well as Sagittarius B, are indicated as lower limit, as their $A$ factor is much higher.}
\label{fig:Arad}
\end{figure}

{  The $\gamma$-ray emissivity of a GMC depends on the ratio of timescales of proton-proton interactions and CR propagation in the cloud. The proposed method of probing the CR density with $\gamma$-rays can be realized provided that CRs freely penetrate the clouds. Although the latter condition cannot be {\it a priori} satisfied (see, e.g. ref.\cite{CesarskyVoelk}), in the case of "passive" clouds embedded in the CR sea, we can safely assume free CR penetration unless the propagation inside the clouds is dramatically slower compared to the diffusion in the ISM \cite{morlino2015cosmic}.
For the typical parameters for GMCs, the timescales of propagation of CRs through the GMCs do not exceed $10^4$~yrs even for low energy ($\leq 10$~GeV) particles, assuming that the diffusion coefficient inside the cloud is similar to the one in the ISM \cite{GabiciFAPB}.  It is then shorter, by two orders magnitude than the confinement time of CRs in the Galaxy ($10^6$--$10^7$ years), as well as the characteristic $pp$ interaction time inside the clouds ($\tau_{pp} \simeq 3 \times 10^5 (n/10^2 \ \rm cm^{-3})^{-1}$~yr).  Otherwise, for $\gamma$ rays, the products of these interactions, we should expect significantly harder spectrum compared to the spectrum of parent protons \cite{GabiciFAPB}.
While this seems quite unlikely to happen in "passive" clouds, the impact of CR propagation effects could be much stronger in the case of clouds located in the vicinity of CR accelerators with the activity timescales much shorter than the CR confinement time in the Galaxy. Therefore, the detection of hard $\gamma$-ray spectra from individual clouds can serve as an indicator of the presence of nearby recent or currently active accelerators, and thus should be excluded from the sample of objects to be used in the extraction of the flux of the CR sea. Apparently, this cannot be done in the case of derivation of the CR sea from the diffuse $\gamma$-ray background, the latter being the superposition of contributions from both \textquotedblleft active" and \textquotedblleft passive" clouds, as well as the intercloud regions.}

\section{The target selection}

{ We analyzed 19 GMCs spread over the Milky Way, from local clouds belonging to the Gould Belt complex to the Central Molecular Zone in the Galactic Center region.  All the analyzed clouds are listed in Table \ref{cloud_tab} and shown in  Fig 3 in Galactocenteric coordinates}. 
\begin{table}[ht!]
\begin{tabular}{|ccccccc|}
\hline
Cloud & $l$  & $b$  & Mass   & d  &  R$_{\rm GC}$ & $A$ \\
      &    (deg)    &     (deg)  &     (10$^5$ M$_\odot$)     &       (kpc)         &        (kpc)   &        \\

\hline\hline

243   & 42.04   & $-$0.36   & { 30 $\pm$ 10}                  & {  7.9 }$\pm$  0.6   & 5.8      & 0.47              \\
418   & 111.45  & 0.79    & { 8 $\pm$ 3 }                & {  4.1 }  $\pm$    0.6  & 10.6     &   0.46         \\
429   & 109.84  & $-$0.29   & { 10 $\pm$ 4}              & {  3.9 } $\pm$    0.6 & 10.3     &   0.67           \\
610  & 142.40  & 1.38     & { 0.3 $\pm$ 0.4}   &     {  0.7}   $\pm$     0.5 &  8.9       &			 0.61 \\
612   & 126.87  & $-$0.66   & { 0.3 $\pm$ 0.5}                   & {  0.6 }    $\pm$  0.5 & 8.7  &    0.83       \\

804   & 328.58  &  0.4     &  { 24 $\pm$ 8}           &  {  5.7} $\pm$   0.6   &      4.6 &   0.75 \\   
876 & 323.61    & 0.22   &  { 60 $\pm$ 18}         &  10.2 $\pm$ 0.4 &  6.0 &  0.58 \\
877   & 333.46  & $-$0.31    & { 13 $\pm$ 4}         & {  3.4 } $\pm$   0.4    & 5.5      & 1.11          \\
900 & 318.07  &  $-$0.21  &  { 47 $\pm$ 14}     &   {  9.8} $\pm$ 0.4  & 6.6 & 0.48 \\
902   & 340.84  & $-$0.30    & { 110 $\pm$ 30}  	& {  12.5 } $\pm$  0.4 &  5.4     &  0.73           \\
933   & 305.49  &  0.11    & { 29 $\pm$ 12}          & {  6.8} $\pm$ 0.9   & 7.1      &     0.63      \\
964   & 345.57  &  0.79    &  {3 $\pm$ 2}           & {  1.9} $\pm$ 0.6   & 6.4      &  0.75 \\     
\hline 
Taurus            &  171.6   & $-$15.8   &    0.11  &   0.141$\pm$ 0.007 & 8.4  &        5.6      \\ 
Lupus             &  338.9   &  16.5   &   0.04   & 0.189 $\pm$ 0.009 & 8.2   &        1.0      \\ 
Orion A           &  209.1   & $-$19.9   &    0.55  & 0.43 $\pm$ 0.02 & 8.4   & 	3.0   \\ 
Cepheus           &  110.7   &  12.6   &    2.13 & 0.92 $\pm$ 0.05 & 8.6     & 2.5     \\ 
MonOB1     &  202.1   &  1.0   &    1.33        & 0.745 $\pm$ 0.03 & 9.1   &   2.4   \\ 
Maddalena         &  216.5   & $-$2.5    &    5.29 &    2.1 $\pm$ 0.1 & 10.1 & 1.2     \\     
\hline
Sgr B & 0.65  & $-$0.05  & 150   &  7.9$\pm$0.8  & 0.1  & 2.3 \\
\hline
\end{tabular}
\caption{Parameters  of the chosen GMCs: Galactic coordinates ($l,b$), masses $M$,  distances from the Earth $d$, 
Galactocentric  distances $R_{\rm GC}$, and the $A$ parameter.  The information about the  distant GMCs  (the upper part of the table) are  from ref.\cite{rice2016uniform},  about distance of nearby clouds (lower part of the table, below the horizontal line) are from ref.\cite{zucker2019large}, {  and the masses of the latter are calculated from the Planck templates}.  For the mass and distance to the Sgr B complex see refs.\cite{yang2015fermi} and \cite{reid2009trigonometric}. 
}

\label{cloud_tab}
\end{table}
The distance of the clouds is determined in different ways: the star-reddening technique can be used, providing very precise estimation, \cite{schlafly2014large,zucker2019large} only for nearby objects. For more distant  objects, unless masers are present, like in the case of the Sgr B complex \cite{reid2009trigonometric}, the kinematic distance method (e.g. \cite{rice2016uniform}) is used. For this reason the uncertainties in the distances of our chosen molecular clouds are quite large.  Nevertheless, it is important to notice that the uncertainties in the distance estimates do not reflect in the  $\gamma$-ray flux, as showed in details in Appendix A.

We differentiate the clouds of our sample into {\it Nearby Clouds}, {\it Rice clouds} and {\it Sgr B}. 


\paragraph{Nearby Clouds.}
In the local ($R \sim $ 200--500 pc) environment, we chose three clouds from the Gould Belt complex.  Two of them, Taurus and Orion A, have been analyzed in previous studies \cite{Neronov2017,yang2014probing} allowing us to cross-check our results.  The third source, Lupus, has not been studied before. In addition, we selected three more clouds that, despite being outside the Gould Belt, are relatively close-by sources, namely Monoceros OB1, Cepheus and Maddalena{  's cloud}.   Note that  recent studies revealed two gaseous structures associated with Cepheus; one at  $\sim$ 300 pc and another at $\sim$800 pc. In this work we considered the most distant structure.  All these GMCs, being close,  have high values of the $A$ parameter and lie several degrees far from the Galactic Plane, where the \fermi sensitivity  becomes better. Therefore they are optimal candidates for this study.

\begin{figure}[h!]
\includegraphics[width=\linewidth]{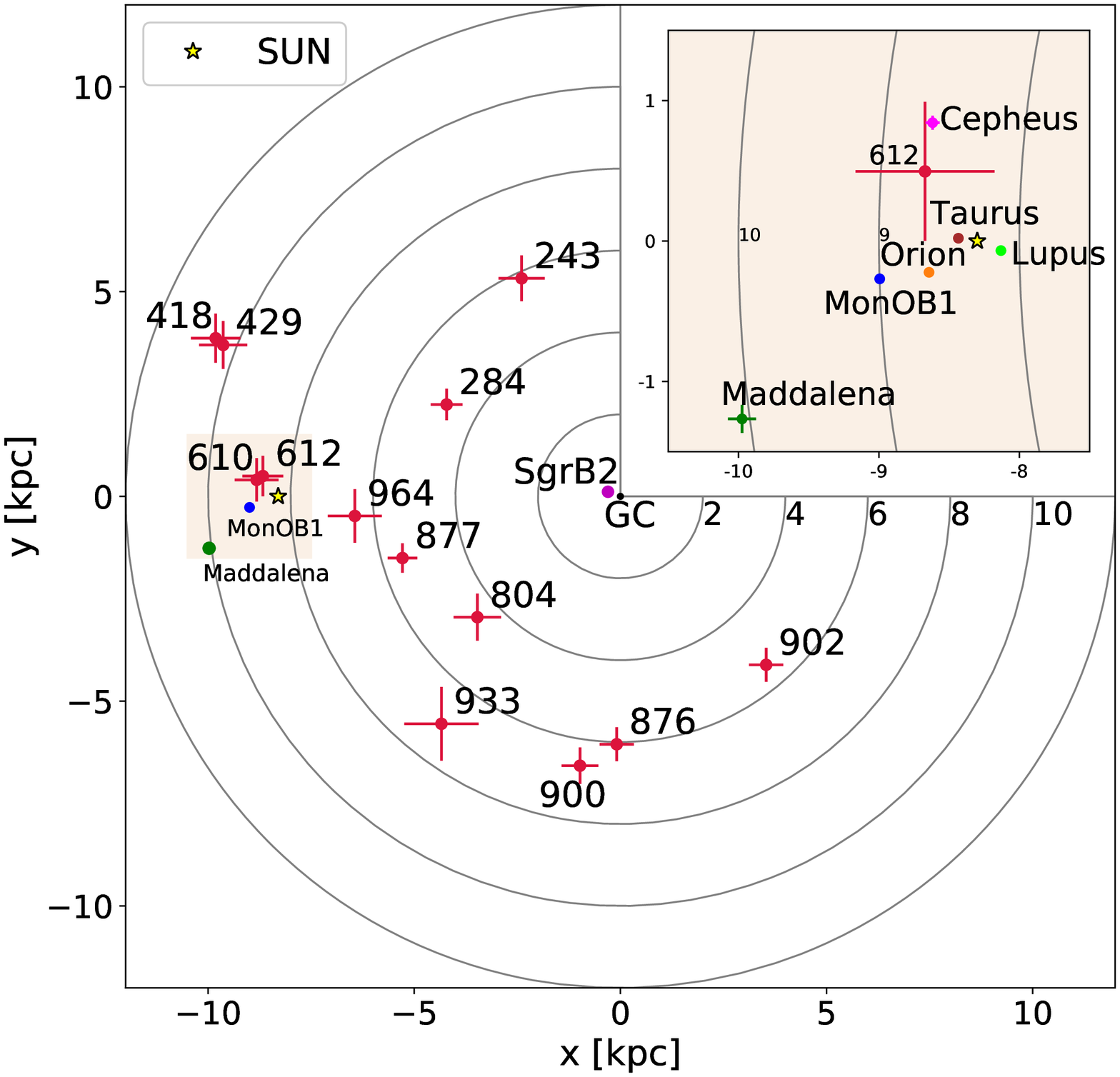}
\caption{Positions  of the selected GMCs  in  the Galactic Plane.   In the upper right panel is shown the zoomed region around  the Sun with a few selected nearby clouds.  The location of the Sgr B complex in the Galactic Center  is also shown. The positions and their relative uncertainties are taken  from refs.\cite{rice2016uniform}, \cite{schlafly2014large} and  \cite{comeron2008lupus}.}.
\label{fig:position}
\end{figure}

\paragraph{Rice clouds.}
The majority of our cloud sample consists of clouds belonging to the Rice et al. catalog \cite{rice2016uniform}. In total, the catalog contains 1064 molecular clouds, distributed  between the longitudes $ 180^\circ > $l$ > 13^\circ$ and $348^\circ > $l$ >180^\circ $, and contained  within the narrow band of latitudes, $-5^\circ < $b$ < 5^\circ $, covering the range of Galactocentric distances  from $\sim$2 to $\sim$15 kpc.   Approximately 4\% of these clouds have masses between  $10^6$  and $10^7$ M$_\odot$ and relatively small angular extensions, $\lesssim1^\circ $. We selected 12 GMCs from this catalog. The selection was based on the following criteria:  (i) the value of the $A$ parameter, (ii) the contribution of the individual clouds to the gas column density  along the given line of sight,  (iii) the presence of resolved gamma-ray sources in the cloud's proximity. 

Remarkably, the very massive GMCs  chosen for the analysis of \fermi $\gamma$-ray data, appeared to be the dominant objects 
in terms of contributions to the total column densities in the corresponding directions. Details of these estimates can be found in the Appendix. For all clouds, the contribution to the column density of the molecular gas varies between 40 \% and 85\%, see Table \ref{tab:gas_percentage}.  This reduces the possible confusion with other  background or foreground clouds,  thus allows us to identify the location of the $\gamma$-radiation sites with the cloud location.

\begin{table}[h!]
\begin{tabular}{|c|c|c|}
\hline
Cloud & $\frac{N_H(cloud)}{N_H(H_2)}$\% &   $\frac{N_H(cloud)}{N_H(H_2+H{\sc I})}$\%\\
\hline
243 & {  53}  & {  30} \\
418 & {  83} & {  45} \\
429 & {  86} & {  31} \\
610 & {  82} & {  29} \\
612 & {  73} & {  19} \\
804 & {  45} & {  26} \\
876 & {  53} & {  27}  \\
877 & {  49} & {  34} \\
900 & {  58} & {  32} \\
902 & {  47} & {  32} \\
933 & {  84} & {  41} \\
964 & {  41} & {  28} \\

\hline 
\end{tabular}
\caption{The relative contributions of the clouds chosen for the gamma-ray analysis, to the  molecular ($H_2$)  and entire  hydrogen  (H$_2$+H{\sc I}) column densities towards the directions of these clouds. }
\label{tab:gas_percentage}
\end{table}

To evaluate the possible confusion with other very bright $\gamma$-ray sources we considered the 3FGL \cite{3fgl} and the HGPS \cite{hgps} catalogs,  and studied the known sources located in the vicinity of our sampled clouds. We discarded the clouds that have many overlapping sources (e.g clouds 151, 190, 842) and the clouds that are in the proximity of strong H.E.S.S. sources (e.g. 269, 292, 897).  The 12 selected clouds do not have identified  nearby sources, but some of them overlap with {  confused} sources. In particular, the clouds 610, 902, 933 do have faint nearby background or foreground sources which, however, after the fits appeared negligible (TS$<$10). The clouds 804, 877 and 964 have strong (TS$>$25) nearby sources. We tested the effect of the existence of these bright sources in the vicinity of our clouds, by accounting for or eliminating them from the background model. This test showed that while the spectral shapes are not affected, the absolute fluxes could vary within 20 \% which  is smaller than the systematic uncertainty. 



\paragraph{The Sagittarius B complex.}
The dense gas complexes in the Galactic Center region, as parts of the Central Molecular Zone (CMZ), provide an opportunity to probe the CR density at a unique location, at the Galactocentric distance $R\approx 0$.1 kpc . The 
so-called Sagittarius B complex Sgr B, which contains the most massive clouds, Sgr B1 and Sgr B2 located in  CMZ, occupies the region  0.4$^\circ$ $\leq$ $l$ $\leq$ 0.9$^\circ$ and $-0.3^\circ$ $\leq$ $b$ $\leq$ 0.2$^\circ$.  The large distance of these objects, $d_{kpc} \sim 7.9 \pm 0.8 $ \cite{reid2009trigonometric}, is compensated by their huge masses. The total mass of this region, derived  from infrared observations,  results in  a value of the $A$ parameter of 2.3. 
     
\section{Data analysis}
\begin{figure}[!ht]
\includegraphics[width=1\linewidth]{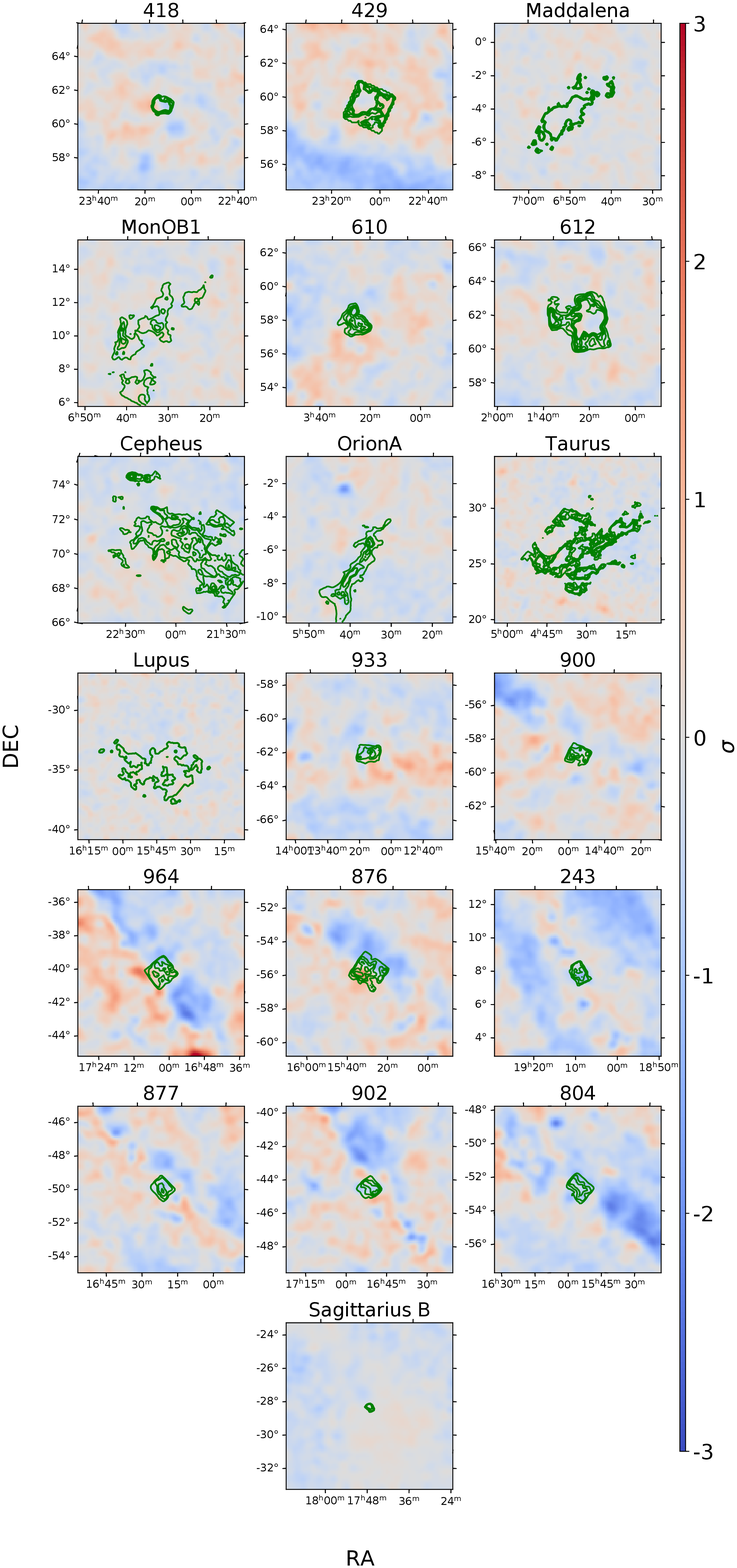}
\caption{Residual maps of the clouds. The green contours show the clouds shape as seen from the CO map, for Rice clouds, and from dust map, for nearby clouds. }
\label{fig:sigma}
\end{figure}

We analyzed 9 years of \fermi{} data, from {MET 239557417} (4th August 2008) to MET 533045411 (22nd November 2017), using the package {fermipy v.0.14.1}. We selected events with an energy larger than 800 MeV as compromise between statistics and good angular resolution. The latter is, in fact,  essential to reduce the effect of the source confusion. We considered Pass 8 data and selected \textquoteleft FRONT+BACK' events (\texttt{evtype=3}) with zenith angles $z \leq z_{max}$=90$^{\circ}$ , to avoid the Earth limb events and imposed \texttt{DATA\_QUAL==1 $\& \&$ LAT\_CONFIG==1}. The considered ROI was a 10$^{\circ}$x10$^{\circ}$ square,  around the cloud centre. As a template for the background, the standard Galactic background model of the \fermi{} collaboration \cite{acero2016development} could not be used since the emission from the molecular clouds is included in the background itself. We generated a customized diffuse interstellar emission model that does not include the emission expected from the selected clouds. For this purpose, we considered the main channels of production of $\gamma$ rays in our selected energy range:   the $\pi^0$-decay radiation, the inverse Compton scattering and the extra-galactic diffuse radiation. We modeled the $\pi^0$ emission from the gas map, by considering different surveys, as discussed below. For  the Inverse Compton component, we used the map ${\rm ^SY^Z10^R30^T150^C2}$ from \texttt{galprop} \cite{galprop} . For the isotropic extra-Galactic component we derived a model by fitting a 30$^{\circ}$ region centered at  $b$=90$^{\circ}$, where the Galactic contribution (pion decay and IC) is minimum. As starting point, we included  the sources from the 3FGL catalog \cite{3fgl} and added, at a latter stage,  new sources that appeared to be significant in the Test Statistic (TS) map. In the likelihood fit, we kept free all diffuse components as well as the normalization of all sources within 3 degrees from the center of the ROI. In Fig. \ref{fig:sigma}, we show the residual maps in terms of $\sigma$, produced after the fitting procedure.  We derived for individual clouds the SED  by fitting each single energy bin with a Power Law (PL) function of index 2 and a normalization parameter left free to vary.  In general, we used  the energy bins corresponding to $\Delta \log E=0.125$, except for some cases when the bins had to be enlarged to provide adequate statistics of counts.

\subsection{Templates}

We constructed the templates for the clouds and for the background $\pi^0$ emission from the radio maps of gas \cite{dame2001milky,bekhti2016hi4pi} or from infrared maps of dust emission \cite{ade2011planck}. From each map, we cut out the cloud and considered the rest of the gas as background. 

For nearby clouds we considered the data from the Planck satellite \footnote{\href{<http://www.esa.int/Our_Activities/Space_Science/Planck>}{http://www.esa.int/Our\_Activities/Space\_Science/Planck}} which provides a full-sky survey of the dust optical depth. We used the maps of the thermal dust optical depth at 353 GHz, as it relates to the interstellar hydrogen linearly \cite{ade2011planck}. The advantage of using dust, is that it maps both atomic and molecular hydrogen. In addition, it is not sensitive to saturation, like CO, and therefore it traces also the so called \textquoteleft dark gas'. The Planck data are bi-dimensional, and can  be used only for isolated clouds, for which any other contribution along the line of sight can be neglected.  

For  Rice clouds,  we need to take into account the background  and foreground gas, 
as  we cannot consider them isolated. We used the 3-dimensional data cube of H{\sc i}from \cite{bekhti2016hi4pi},  and the cube of CO from \cite{dame2001milky} as tracer of molecular hydrogen.  For each cloud, we considered the position ($l_0$,$b_0$,$v_0$), the velocity dispersion $\sigma_v$, and the observed extension $\sigma_r$ as given in the catalog of Rice et al. We consider a cubic box centered at the center of the cloud, with the width $\Delta v$ = 2$\sqrt{2\ln2}\sigma_v$  and the side $2R$; $R$ is the radius of the H$_2$ region inferred from $\sigma_r$.  As explained in \cite{rice2016uniform} and \cite{rosolowsky2008structural}, because of the different gas density profiles, the radius of the CO emitting region $\sigma_r$ is generally smaller than the radius $R$ of the H$_2$ region. Following \cite{rosolowsky2008structural}, we assumed R=$\eta\sigma_r$  with $\eta=1.91$. 
The remaining H$_2$ is taken as background. We considered H{\sc i}as background, but not as a signal. 

For the Sgr~B complex,  we followed the methodology of the previous analysis as described in \cite{yang2015fermi} and updated the analysis by using 10 years of Pass 8 data. We note that it is impossible to perform a kinetic separation towards the Galactic center region, and that, due to the high opacity, CO observations may significantly underestimate the gas density in this region. 
Therefore we used the Planck opacity map \cite{ade2011planck} to trace the gas in this region. We adopted the same data preparation procedure as in other regions, and considered the $0.5^{\circ}\times0.5^{\circ}$ box to define the Sgr~B region as $0.4^{\circ}<l<0.9^{\circ},-0.3^{\circ}<b<0.2^{\circ}$. We used the dust opacity map in this box as our source template and considered the other regions of the map as background. 

\section{Results}


\subsection{Spectral Energy Distribution}
\begin{figure}[!ht]
\includegraphics[width=9.5 cm, height=12.5 cm]{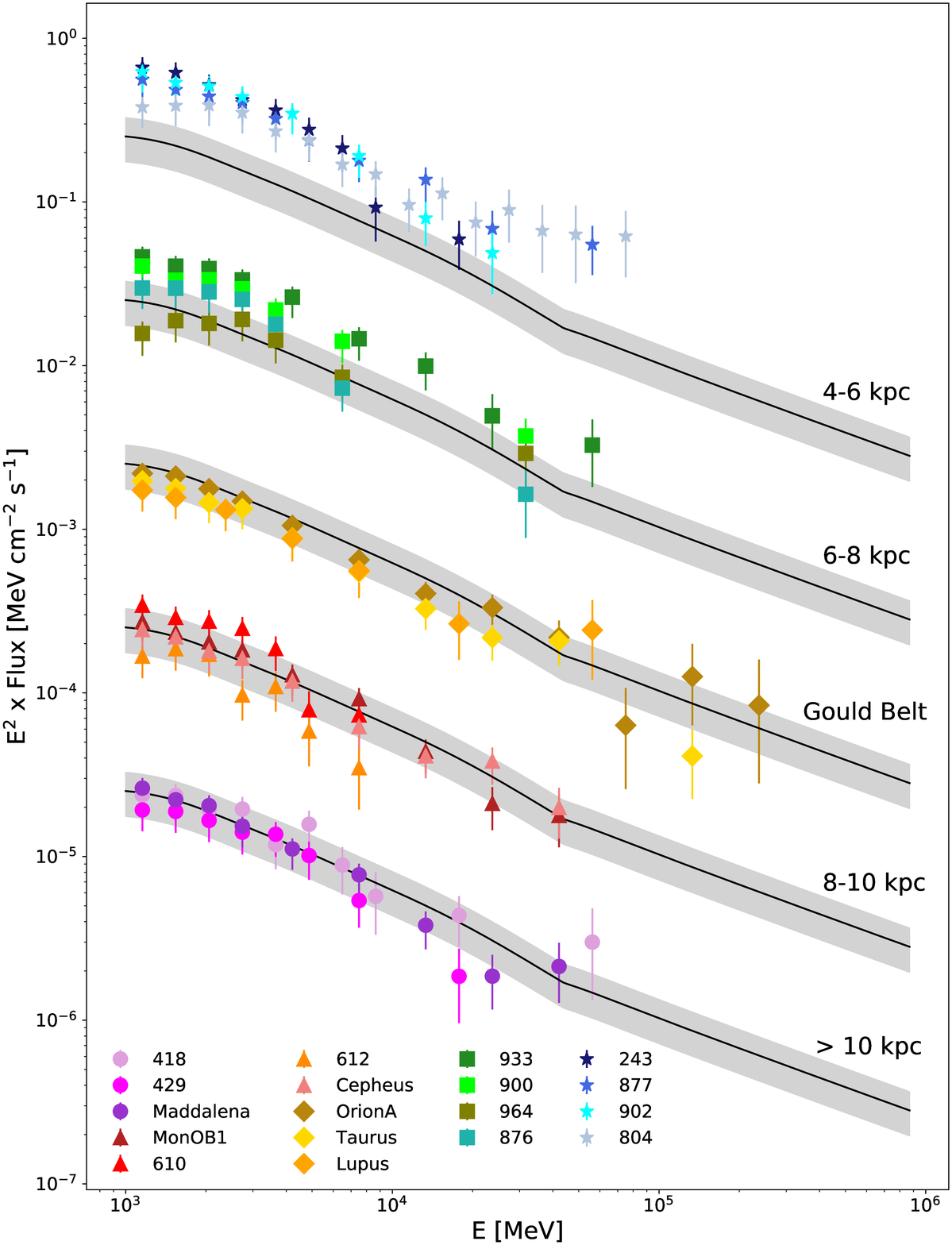}
\caption{\footnotesize{Spectral Energy Distributions of GMCs.  From the bottom to the top: clouds within the Galactocentric distances 10--12~kpc; 8--10~kpc, Gould Belt complex,
6--8~kpc,  4--6 kpc. SEDs are normalized to $A=1$. For the better visibility,  the fluxes corresponding these zones are separated from each other by scaling them by the factors $\rm 10^n$ with n=1,2,3 and 4. The dashed black lines, for each set of GMCs, are the expected SED of a cloud with $A=1$  calculated for the parent CR proton spectrum reported by the AMS collaboration  \cite{ams_proton},  and using the differential cross-section parametrization from  \cite{kafexhiu14}.  The parameter that takes into account the contribution of nuclei to the $\gamma$-ray production, $\xi_N=1.8$,   corresponds to the  standard compositions of the interstellar {  medium} and CRs. The shaded gray area indicates the  30\% uncertainty in M$_5$/d$^2_{kpc}$, due to the CO-to-H$_2$ conversion factor. }}
\label{fig:sed_compare}
\end{figure}

In Fig. \ref{fig:sed_compare} we show the SED of the 19 chosen GMCs. In the same figure, we show the $\gamma$-ray fluxes  expected from  interactions of the sea of CRs with the clouds.
For convenience, when comparing the fluxes from different clouds, the SEDs are normalized to $A=1$.  
The statistical significance of all spectral points, shown in Fig. \ref{fig:sed_compare}, exceeds  TS=5. 
The interactions of the CR sea with the clouds set the minimum level of $\gamma$-ray fluxes from individual clouds.
This component of radiation is calculated for the locally measured fluxes of CR protons reported by the AMS collaboration \cite{ams_proton} as in Eq.\ref{eq:expflux}.  The dashed gray zones indicate the flux uncertainties introduced by $\approx 30$\% uncertainty in the CO-to-H$_2$ conversion factor  $X_{CO}$ that affects the estimation of $A$. 

\subsection{Derivation of Cosmic Ray density}
From the SEDs of the clouds shown in Fig.\ref{fig:sed_compare}, we 
can extract the energy density of the parent CR protons.  The differential $\gamma$-ray flux given by Eq.(\ref{eq:expflux}) can be expressed via the CR density,  $\rho_{CR}= \frac{4\pi}{c} F(E_p)$: 
\begin{equation}
F_\gamma(E_\gamma) \propto \frac{M_5}{d^2_{kpc}} \int dE_p \frac{d\sigma}{dE_\gamma} \frac{4\pi}{c} \rho_{CR} .
\label{eq:rhoCR}
\end{equation}
To calculate $\rho_{CR}$, we used the \texttt{naima} \cite{naima} software package. Compared to the average density of interstellar medium,  the gas density in GMCs is very high. Therefore, in the energy interval of interest, the $\gamma$-ray production via collisions of CR nuclei with the ambient gas well dominates over other $\gamma$-ray production channels.

The \texttt{naima} package uses a modified presentation of Eq. (\ref{eq:rhoCR}):
\begin{align*}
    F_\gamma(E_\gamma) &= \frac{<n_H>}{d^2} \int dE_p \frac{d\sigma}{dE_\gamma} \int_V dV \frac{4\pi}{c} \frac{dN_p}{dE dV} \\
    &= \frac{<n_H>}{d^2} \int dE_p \frac{d\sigma}{dE_\gamma}\frac{4\pi}{c} \frac{dN_p}{dE} 
\end{align*}
To avoid large uncertainties, we considered only the $\gamma$-ray fluxes above $\sim 1 $~GeV  which trace $\gtrsim$ 10 GeV CRs. We assumed the initial form of the CR distribution to be a simple power law, $dN_p/dE = F_0 (E/E_0)^{-\alpha} $, and derived the parameters $F_0$ and $\alpha$ from the distributions of $\gamma$ rays and the ambient gas. Initially we set $<n_H>$=1 cm$^{-3}$ and $d_{kpc}$=1, so that the normalization is $F'_0=\frac{<n_H>}{d^2_{kpc}} F_0 $. Then, 
the normalized CR density is linked to $F'_0$ through the parameter A: 



$$\rho_{0,CR} = \frac{F_0}{V}= 
\frac{m_p}{10^{5} M_\odot} A^{-1} F'_0 \ .$$ 

In this way, the systematic error on the estimation of $F_0$ is reduced to the uncertainty of the parameter  $A= M_5/d^2_{kpc}$ caused basically by the uncertainty  related to the conversion factor $X_{CO}$. The derived  values of $\rho_0$ and $\alpha$ for all clouds are listed  in Table \ref{tab:par_cr} . In Fig. \ref{fig:cr_radial}, we show these values as a function of the Galactocentric distance, $R_{gal}$. { We compared the derived values to the local values for cosmic proton as measured by AMS02. Since AMS02 data are not well represented by a single power law distribution to guarantee a fair comparison we fitted the data in restricted energy intervals. In the interval of our interest i.e. 20--200 GeV, the best Power Law index resulted to be $\alpha=2.8$. In Fig. \ref{fig:cr_radial} the yellow band represents the [2.75,2.85] interval, which takes into account of the systematic variation of the spectral index due to the different range of the fitted points (e.g. Taurus and Lupus have different spectral indices because the spectrum of Taurus extends at higher energies, and therefore it is less influenced by the flattening of the spectrum at few GeVs  )   }

\begin{table}[ht]
    \centering
    \begin{tabular}{|ccc|}
    \hline
 {Cloud} &$\mathbf{\rho_{0,CR}}$  & $\mathbf{\alpha }$  \\
 & [$10^{-12}$ GeV$^{-1}$ cm$^{-3}$ ]& \\
\hline
418 & 1.74 $\pm$ 0.5 & 2.69 $\pm$ 0.05 \\
429 & 1.44 $\pm$ 0.4 & 2.74 $\pm$ 0.05 \\
Maddalena & 1.79 $\pm$ 0.6 & 2.93 $\pm$ 0.06 \\
MonOB1 & 1.93 $\pm$ 0.6 & 2.99 $\pm$ 0.06 \\
610 & 2.34 $\pm$ 1.1 & 2.79 $\pm$ 0.04 \\
612 & 1.29 $\pm$ 0.4 & 2.8 $\pm$ 0.09 \\
Cepheus & 1.68 $\pm$ 0.5 & 2.87 $\pm$ 0.07 \\
OrionA & 1.55 $\pm$ 0.5 & 2.83 $\pm$ 0.05 \\
Taurus & 1.43 $\pm$ 0.5 & 2.89 $\pm$ 0.05 \\
Lupus & 1.09 $\pm$ 0.4 & 2.74 $\pm$ 0.1 \\
933 & 3.19 $\pm$ 1.1 & 2.69 $\pm$ 0.02 \\
900 & 2.71 $\pm$ 0.8 & 2.74 $\pm$ 0.03 \\
964 & 1.33 $\pm$ 0.4 & 2.56 $\pm$ 0.04 \\
876 & 2.26 $\pm$ 0.7 & 2.82 $\pm$ 0.03 \\
243 & 4.78 $\pm$ 1.4 & 2.86 $\pm$ 0.03 \\
877 & 3.87 $\pm$ 1.2 & 2.69 $\pm$ 0.02 \\
902 & 4.41 $\pm$ 1.3 & 2.74 $\pm$ 0.02 \\
804 & 2.98 $\pm$ 0.9 & 2.61 $\pm$ 0.02 \\
Sgr B  & 0.98 $\pm$ 0.06& 2.80 $\pm$ 0.03 \\
{  AMS02} &  {1.12} &  {2.8}   \footnote{From fitting of experimental points on the energy range 20 GeV -- 200 GeV } \\
\hline
    \end{tabular}
   \caption{The spectral indices and  CR proton densities at 10 GeV  derived from the $\gamma$-ray and CO data at the locations of  clouds indicated in Fig. \ref{fig:position}. Errors on the normalization result from the sum in quadrature of the statistical error deriving from the fit and the 30\% uncertainty on the $A$ parameter.}
\label{tab:par_cr}
\end{table}

\section{Discussion}

\begin{figure}[h!]
\includegraphics[width=1.\linewidth]{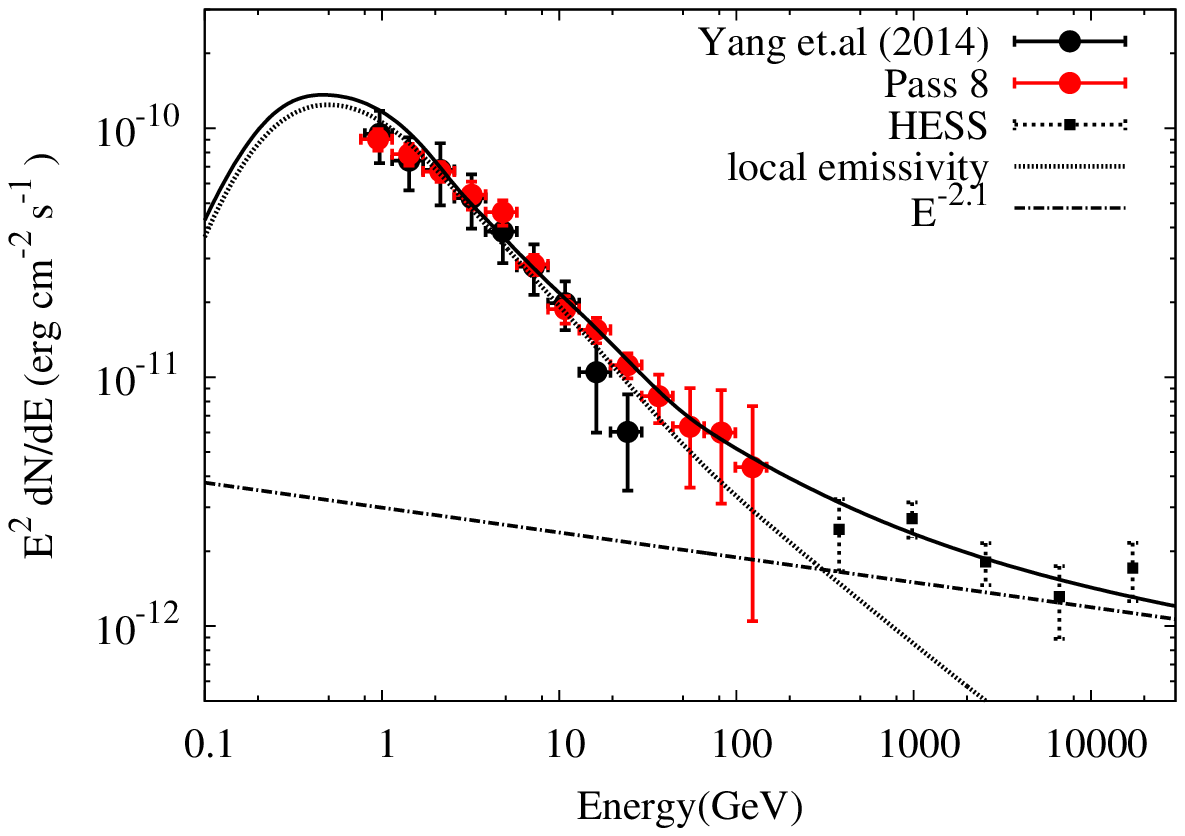}
\caption{$\gamma$-ray fluxes from the Sgr B complex. The red circles are from the 10-years Pass 8 data derived in this work, the black circles are from \cite{yang2015fermi}. The black squares represent the diffuse $\gamma$-ray flux reported by the H.E.S.S. collaboration \cite{hessgc06}. 
The dotted line represents the predicted gamma-ray spectrum under the assumption that the CR spectrum in the Sgr~B region is the same as the local CR spectrum measured by AMS02. 
The dot-dashed line is a power-law component with the photon index of $-2.1$.  In superposition with the \textquotedblleft CR sea"  linked component (solid curve), it  
explains the hardening of the $\gamma$-ray spectrum above 10~GeV and almost entirely the flux at  TeV energies.}
\label{fig:sgrb}
\end{figure}

The results presented in Table III and Fig.\ref{fig:sed_compare} allow a first robust conclusions for {\it all} clouds with Galactocentric distances larger than 8~kpc, independently of their location in the Galaxy.  The results are in good agreement with theoretical predictions assuming that  $\gamma$ rays are produced at interactions of CRs with the gas inside the clouds and that the flux and spectral shape of CRs embedded in the clouds are close to the locally measured CR flux as reported by the AMS collaboration \cite{ams_proton}.
From Fig. \ref{fig:sed_compare} we can see good agreement between the observed   $\gamma$-ray fluxes from three regions representing the Gould Belt complex, the 8--10~kpc ring and the periphery beyond $\geq 10$~kpc. While the data relevant to the Gould Belt clouds could be interpreted as a  result of the dominant contribution by local accelerators to the measured CR flux,  the $\gamma$-ray data from other GMCs, in particular, Maddalena and the clouds $\#$418 and $\#$429, that lie further away ($\gtrsim$ 1 kpc), exclude this scenario.  {  In addition, the same nominal flux   (i.e. a $\gamma$-ray flux expected from the AMS type CR flux) is observed from the cloud \#964 located in the inner Galaxy, at a distance of 6 kpc from the GC. } The constancy of the derived  densities and the spectral indices of CRs tell us that, most likely,  we deal with the sea of CRs. Surprisingly,  the same level of CR density has been found by Yang et al. \cite{yang2015fermi} and reconfirmed by Pass 8 data analysis also for the Galactic Center region. 
The new analysis of $\gamma$ rays with almost doubled photon statistics confirms the previous conclusion about the low CR flux \cite{yang2015fermi} in the CMZ. The  $\gamma$-ray SED  from this region is shown in  Fig.\ref{fig:sgrb}. Below 10 GeV,  one can see a very good agreement of the previously reported fluxes \cite{yang2015fermi}. However, above 10 GeV, the new analysis reveals a significant hardening which is naturally explained by the presence of the diffuse component of very high energy $\gamma$ rays discovered from the same region with the H.E.S.S. telescopes \cite{hessgc06}.  Because of the harder spectrum, the contribution of this component below 10 GeV becomes less than 10\%.  The parent particle population responsible for this component is provided, most likely, by the PeVatron located within the central 10 pc region of the GC \cite{abramowski2016acceleration}. 
Meanwhile, the low energy component is perfectly explained by the sea of Galactic CRs. This is a rather unexpected result, given the presence of several potentially powerful CR accelerators linked to the central supermassive black hole and the high starburst rate in this area.  A plausible explanation of this  result could be the effective escape of  low-energy CRs from  the inner parts of the GC, e.g., due to the fast convection, before they could propagate to large distances and  approach the Sgr~B  complex. 

{  
An alternative interpretation of the enhanced, hard spectrum CR component derived from the H.E.S.S. and \fermi observations of CMZ was proposed by Gaggero et al.\cite{Gaggero}.  They argued that the enhanced CR flux towards the GC could be explained by a  specific CR transport model assuming a position-dependent diffusion coefficient.  However, this model cannot address the $1/r$ distribution of CR density above 10 TeV as derived from the H.E.S.S. observations within the CMZ. The authors admit \cite{Gaggero} the need of the existence of CR source(s) at the heart of GC. Moreover, the very low density of $\leq 1$~TeV CRs derived from \fermi $\gamma$-ray observations at the position of Sgr B complex (see Fig.\ref{fig:sgrb})  excludes the interpretation of ref.\cite{Gaggero}.
}


\begin{figure}[!ht]
    \centering
    \includegraphics[width=1\linewidth]{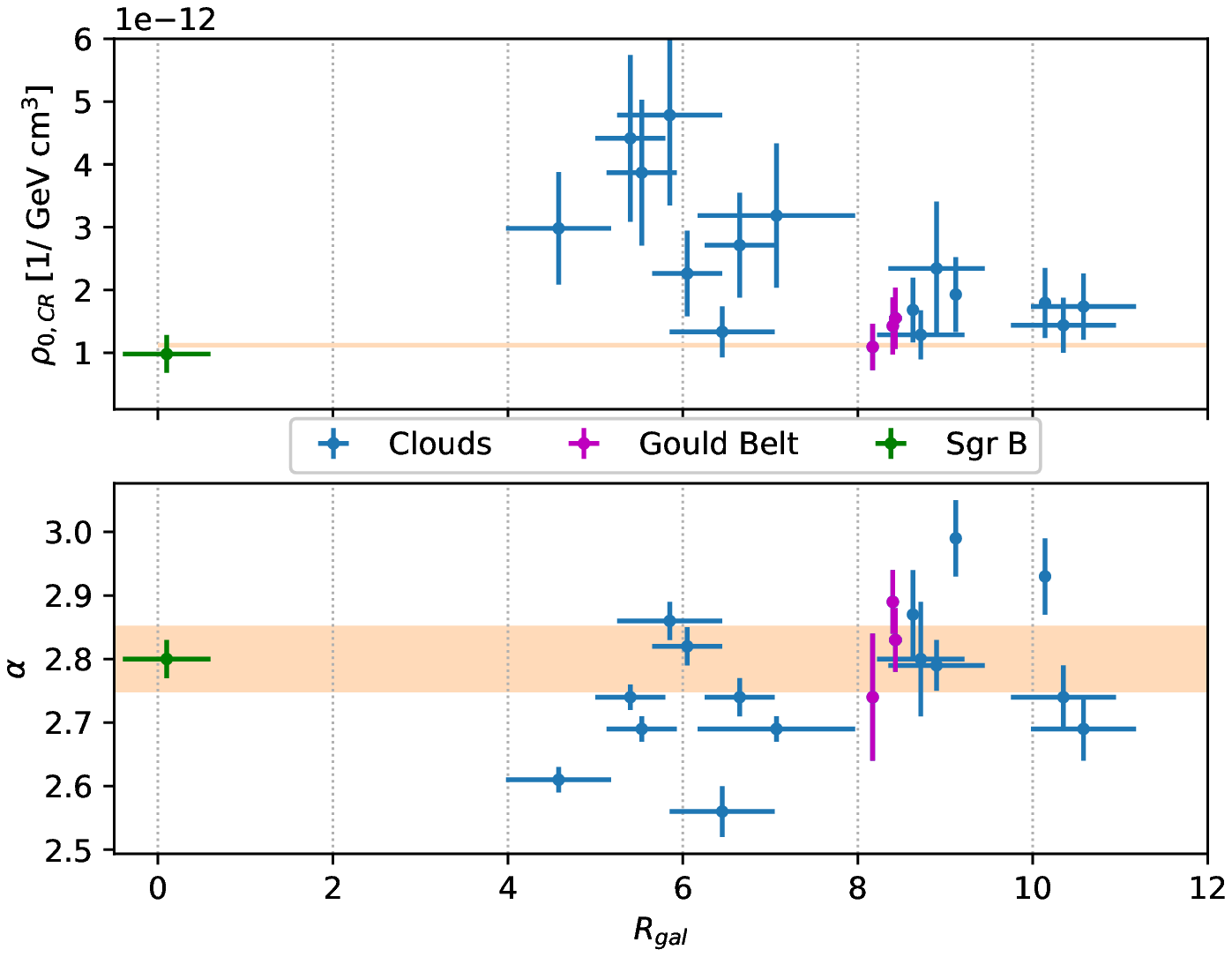}
    \caption{Cosmic Ray density parameters plotted against the galactocentric distance of the clouds: normalization $\rho_{0,CR}$ evaluated at 10 GeV, and spectral index $\alpha$. }
    \label{fig:cr_radial}
\end{figure}

The only noticeable deviation from the homogeneous CR sea in the Milky Way, regarding both the level and the spectral shape, is detected in the clouds 
located in the {  inner Galaxy, within 4 and 6 kpc.} 
{  Some of the clouds there}  show enhanced,  by a factor of 2 or 3,  $\gamma$-ray fluxes,  as well as systematically harder energy spectra compared to  $\gamma$ rays induced by the sea of CRs (see Fig.\ref{fig:sed_compare} and Fig. \ref{fig:cr_radial}). This enhancement was already found in the diffuse $\gamma$ radiation of the Galactic Disk  \cite{Stecker75,bloemen1986radial,yang16,acero2016development, pothast2018progressive}.  

However, {  only} the $\gamma$-ray emission from individual clouds can provide {\it direct} information about the CR density in the specific regions,
which can tell us whether the enhanced $\gamma$-ray emission of the 4--6 kpc ring is a result of a global variation of the level of the CR sea on large (kpc)  scales or it is  caused by an additional component of radiation on top of the homogeneous CR sea.  From our results we do see indication  of fluctuations both in the flux and in the spectral index of $\gamma$ rays in the the 4--6 kpc clouds. {  In addition one can observe stronger fluctuations  for  the clouds located at the galactocentric distances of  $\sim$ 6 kpc, namely \#964, \#876 and \#243   (see Figs. \ref{fig:sed_compare} and  \ref{fig:cr_radial}). Thus } we give  preference to the second scenario, even though the variations do not significantly exceed the systematic uncertainties.  

On the other hand,  a detection of  a \textquotedblleft nominal" $\gamma$-ray flux  even  from  a single cloud in the 4--6 kpc ring, would imply that the variation of the CR density in the ring is a result of presence of an active or recent accelerator in the proximity of the cloud. While we do not  have an appealing case for the 4--6 kpc ring, $\gamma$ radiation from at least one cloud within the 
6--8 kpc ring reveals CR density which is quite close to the one measured by AMS. Although this cloud formally enters the next, \textquotedblleft 6--8~kpc"  zone, its distance to the GC is compatible with 6 kpc,  even considering the distance uncertainties. 
Note that one should not expect a sharp edge of the  \textquotedblleft4--6~kpc" ring,  especially given that the stellar volume emissivity at 6.5 kpc drops only by a factor of two compared to the maximum at 4.5~kpc \cite{popescu2017radiation}. Therefore the \textquotedblleft nominal" $\gamma$-ray flux could imply a minimum, close to the local (AMS) CR density in this part of the ring. What concerns the low level of CR density,  an obvious reason could be the absence of a nearby active CR accelerator. 
{  Also, the steep energy spectrum argues in favour of free penetration of CRs into the clouds \cite{GabiciFAPB}}. 

{  Based on the above arguments we propose} that the  radial variations of both the density and energy spectrum of CRs   derived from the analysis of  the diffuse $\gamma$-ray emission of the Galactic Disk \cite{yang16, acero2016development},  is not related to the  the global  CR sea, but has a more local character caused by the large number of cosmic accelerators in the   4--6~kpc  ring and correspondingly higher CR density in their neighbourhoods.  This seems a direct  consequence of the fact that most of the active star forming regions, and therefore, the potential particle accelerators, are located  within the 4--6~kpc ring. The scales of regions with enhanced CR density depends on the strength and the age of the accelerator.  For SNRs with $10^{51} \ \rm erg$  mechanical energy release,   the  CR density in the surrounding ISM can exceed the level of the CR sea up to 100 pc from the accelerator \cite{AA96}.   In the case of young stellar clusters with an overall mechanical luminosity of stellar winds $10^{38-39} \ \rm erg/s$, the regions with enhanced CRs density  can extend well beyond 100~pc \citep{YSC}. 
Since both types of accelerators are tightly linked to the  star-forming regions,  they are often surrounded by GMCs. The presence of dense gaseous regions close to particle accelerators creates favourable conditions for effective production of $\gamma$ rays \cite{AA96} .

\section{Summary}

{  The results obtained in this study unveil a homogeneous sea of Galactic CRs with a constant density and spectral shape close to the flux of directly measured CRs. This concerns the most of the Galactic Disk, except for, the galactocentric 4--6~kpc ring region. Furthermore, we found a hint for a variation of the CR density in different locations within the same galactocentric ring. We argue that if this is a real effect, but not a result of statistical fluctuations, the enhanced density and harder energy spectra of CRs are likely to be caused by the location of active CR accelerators in the vicinity of the GMCs. Thus, the level and the energy spectrum of the CR sea in the 4--6 kpc ring could be the same as in the other parts of the Milky Way.

An in-depth study of this issue requires significantly larger statistics of $\gamma$-ray emitting clouds and denser coverage of  Galactocentric distances.  Unfortunately,  the lack of projects for a new space-based $\gamma$-ray telescope with significantly improved, compared to the \fermi{} sensitivity at GeV energies, does not give us much hope, at least for the near future, for a significant increase of the number of clouds resolved in $\gamma$ rays.   The prospects are more promising at higher energies. In particular, the sensitivity of the future Cherenkov Telescope Array (CTA) seems to be sufficient for detection of a few \textquotedblleft passive" clouds characterized by the parameter $A \geq 1$ in the energy region between 100 GeV and 1 TeV. Moreover, the detection threshold of CTA 
for clouds characterized by enhanced CR density and harder spectra could be as small as $A\sim 0.1$. This would dramatically increase the number of the potentially detectable clouds providing a denser mapping of CRs. The extension of studies to energies up to 100 TeV and beyond requires more sensitive detectors like LHAASO or SGSO. }

--





%

%% file: Appendix.tex
\section{Mass Determination}
To have a consistent estimation of the mass, we derived it from our cut-out template, as the mass given in the catalog by \cite{rice2016uniform}
might refer to a slightly different shape.  Following Rosolowsky \cite{rosolowsky2008structural}, the mass and consequently the parameter $A$=M$_5$/d$^2_{kpc}$ is computed from the column density, derived from the CO emission:
$$ M =  2 m_p X_{CO} \sum_{i,j} W_{ij}(l_i, b_i) \Delta l \Delta b \bigg(\frac{\pi}{180}\bigg)^2 d^2 \newline $$ 
where $ W_{ij}(l_i, b_i) \equiv \int^{v_{max}}_{v_{min}} T_{CO}(l,b,v) dv $ is the integrated brightness temperature of CO for every pixel. We can notice that the mass is directly related to the distance $d^2$, since it links the physical extension to the observed angular one. This allows us to have an estimation of the $A$ factor directly from the CO data, that does not depend directly on the estimation of the distance and the mass:
$$A= \frac{M_5}{d^2_{kpc}} \approx  32 \sum_{i,j} W_{ij}(l_i, b_i) \Delta l \Delta b \bigg(\frac{\pi}{180}\bigg)^2  $$
As a consequence the  uncertainties on the mass and on the distance cancel out and the only uncertainty on $A$ comes from the factor  $X_{CO}$, that is considered to be of 30\% as suggested by Bolatto \cite{bolatto2013co}.  

The Helium fraction of the ISM here is not taken into account in the calculation of the mass, differently from what is done in \cite{rice2016uniform} where the authors include it as a further factor 1.36. Helium contribution in $\gamma$-ray emission from pp interaction is already accounted in the nuclear enhancement factor, that we assumed to be 1.8 as in \cite{kafexhiu14}.  

\section{Gas distribution}
\begin{figure*}
\includegraphics[width= 1\linewidth]{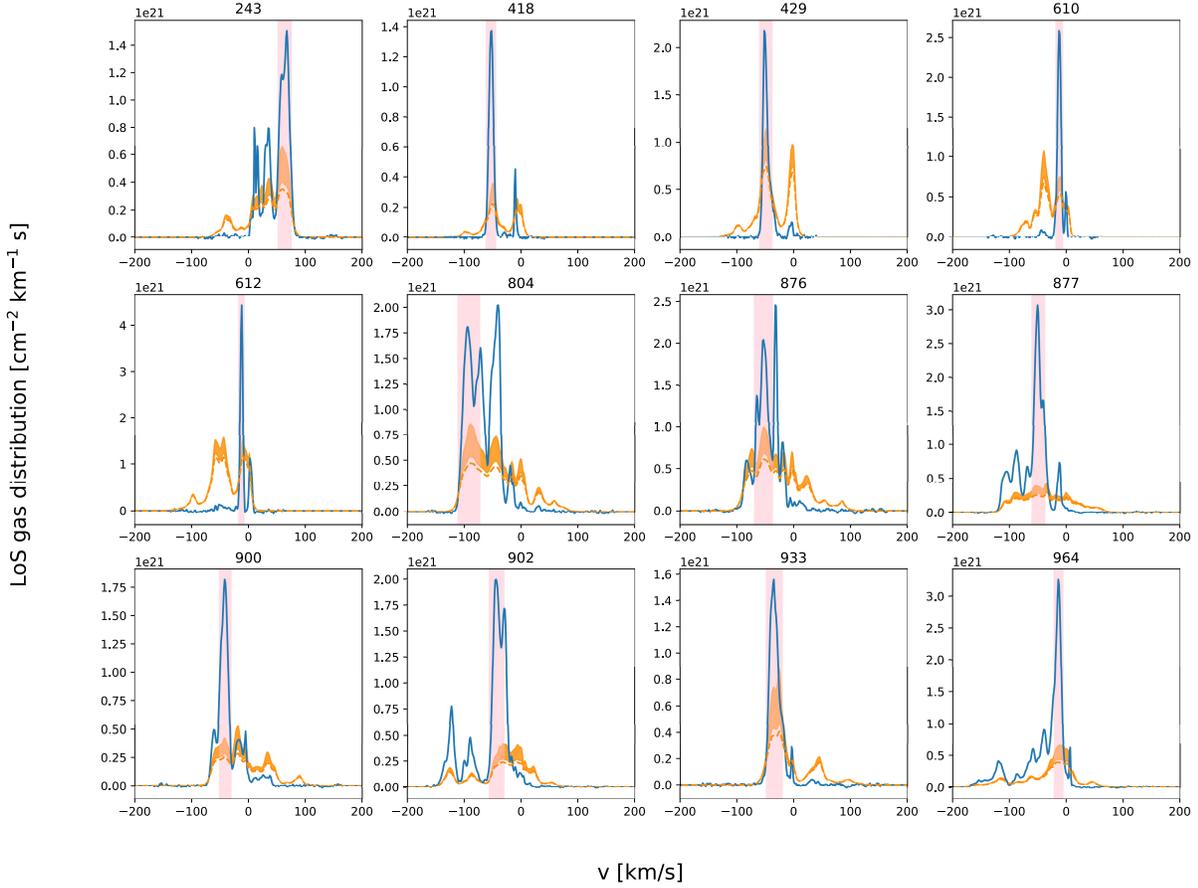}
\caption{Gas distribution along the line of sight in the regions of the analyzed MCs. The pink area delimits the velocity-range of the cloud. Blue lines  represent the distribution of molecular hydrogen for which we assumed X$_{CO}=2\times 10^{20}$ cm$^{-2}$ K$^{-1}$ km$^{-1}$ s.  The orange dashed lines trace the HI distribution, assuming optically thin emission with a conversion factor of X$_{HI}=1.8\times 10^{18}$ cm$^{-2}$ K$^{-1}$ km$^{-1}$ s. The orange area takes into account the variation of density while assuming a spin temperature correction with 150 K $\leq T_s \leq 500 $ K .}
\label{fig:gasprofile}
\end{figure*}

Differently from its atomic counterpart, the molecular hydrogen (H$_2$) is not uniformly distributed in the Galactic Disk. 
The molecular gas in the interstellar medium (ISM) rather  tends to concentrate in  dense massive  clouds. 
Correspondingly, the column density  of the gas in given direction is composed of contributions from most massive clouds. 


The best tracer of molecular hydrogen in optically thin regions is the mm emission of  $^{12}$CO(1$\rightarrow$0). The column density 
in the given direction is determined as:

\begin{equation}
n(l,b) \ [cm^{-2}] = X_{\mathrm{CO}} \int \mathrm{d}v \ T_b(l,b,v) \ [\mathrm{km \ s^{-1} \ K}]   
\end{equation} 
where, following  \cite{bolatto2013co}, the CO-to-H$_2$ conversion factor, $X_{\mathrm{CO}}$, is set to $X_{CO}= 2 \times 10^{20}$ cm$^{-2}$ K$^{-1}$ km$^{-1}s$. In this paper, we use the  CO data from \cite{dame2001milky}.

In the analysis, the atomic hydrogen is considered only as background gas  since it is difficult  to isolate the HI in the cloud from the rest of the column density. For the atomic counterpart we considered the 21cm emission line data-cube from the HI4PI collaboration \cite{bekhti2016hi4pi}.  As for the molecular gas, the HI density is directly proportional to the brightness temperature of the 21-cm line. In this case, for the conversion factor we use the value from \cite{bekhti2016hi4pi}: $X_{HI}=1.83 \times 10^{18}$ cm$^{-2}$ K$^{-1}$ km$^{-1}$ s.  This conversion factor is derived from the assumption of optically thin emission and must be corrected for possible absorption caused by cooler components. Following for example\cite{sofue2017optical}, the corrected column density is calculated as: 
\begin{equation}
n_{HI}(l,b) [cm^{-2}]= -X_{HI} T_s\int dv \ln \big(1-\frac{T_b}{T_s-T_0}\big)
\label{eq:nhicorr}
\end{equation}
where $T_s$  is the average spin temperature of the interstellar HI  \cite{dickey2009outer}. $T_s$ is an effective parameter that describes the mixture of warm and cold neutral components: \begin{equation}
T_s= T_{CNM} \frac{n_{w}+n_{c}}{n_{c}} 
\label{eq:TS}
\end{equation} 
The value of $T_s$ can be measured in presence of strong continuum sources in the background as described, for example, in \cite{walsh2015survey}. These measurements are available only for specific regions of the Milky Way, for other locations we need to rely on an educated guess. For our case we considered that the maximum measured brightness temperature in these clouds is 143 K, so we assumed, as lower limit, $T_{s,min}$=150 K; as a upper limit we assumed  $T_{s,max}$=500 K, that results from eq. \ref{eq:TS} by taking $n_w=$90\% of warm neutral medium and  $T_{CNM}=$50~K \cite{heiles2003millennium}. 

In Fig. \ref{fig:gasprofile} we show the derived H$_2$ and HI distributions by integrating $T_b(l,b,v)$ in the regions that coincide with the clouds spatial templates, defined as explained in the main text. The pink area represents the cloud extension along the line of sight.  


\subsection{Systematic uncertainties}\label{sec:sys}
We considered the following sources of systematic errors for the spectral points:
\paragraph{New sources.} At the moment of writing the 4FGL catalog has not been publicly released, so we made use of the 3FGL catalog of $\gamma$-ray sources. The 3FGL catalog is based on 4 years of data, so we expect to reveal several new sources.  
In few cases some sources appeared at the edges of our analyzed clouds. {   We evaluated the uncertainties deriving from the new sources with this procedure: we firstly fitted the data with a model that included (besides the galactic and extragalactic emission) only the known point sources from the 3FGL catalog and derived the SED for the Cloud with this model. Then we identified as new sources all the significant (TS$>$20) residuals  in the R.O.I. and included them in the model. We refitted until the TS map was clear from significant sources (fig 4). We checked also the residual maps to be sure that they were ranging from $-$3 to 3 sigma.  With this new model we derived a new SED for the Clouds. The difference between the two SEDs was then considered as an additional uncertainty.} 
The difference from the two sets of spectral points is estimated to be at worst $\sim$ 20 \% for what concerns the normalization and does not affect the slope. This uncertainty is included in the systematic errors for each cloud.

\paragraph{Radial dependent CR distribution.} Studies of the diffuse $\gamma$-ray emission showed that the cosmic ray spectra have different shapes at different distances from the GC. In particular \cite{yang16, acero2016development} show that the CR density increases and spectrum hardens towards the inner Galaxy. In our analysis we assumed the CR radial distribution to be uniform in the considered ROI.  We evaluated the effect of the radial dependence of the background \cite{yang16} by performing an equivalent analysis but assuming a radial dependent background model. We divided the gas in 6 Galactocentric rings and assigned to each of them the corresponding normalization and spectral index of CRs derived in \cite{yang16}. The difference between the SEDs derived with the two distinct backgrounds is lower than 10\% in terms of normalization. 

\paragraph{Optically thin H{\sc I}.} As discussed before, the map of  H{\sc I} from the  H{\sc I}4PI collaboration does not have any assumption on the optical thickness of the gas. We can't assume that the H{\sc I}, especially in the galactic disk, is optically thin as this could lead to an underestimation of the background. We corrected the computed column density by using equation \ref{eq:nhicorr} with different values of $T_s$ and tested the influence of this variation by performing an independent analysis on the same data. Even if for all selected clouds the contribution of the molecular gas is dominant over the atomic, in terms of column density, the resulting spectra seem to be affected by the choice of the HI background by a factor that could vary from 20 to 30 \%. 


All these contributions were then summed in quadrature to obtain the final systematic uncertainty.


\subsection{Counts profile}
\begin{figure}[h]
\subfigure[]{ \includegraphics[width=1 \linewidth]{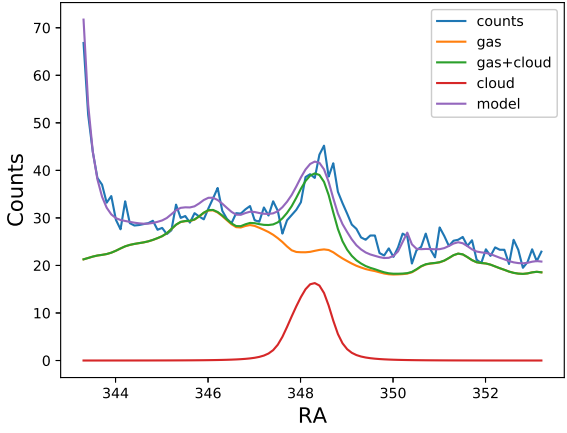}}
\subfigure[]{\includegraphics[width=1 \linewidth]{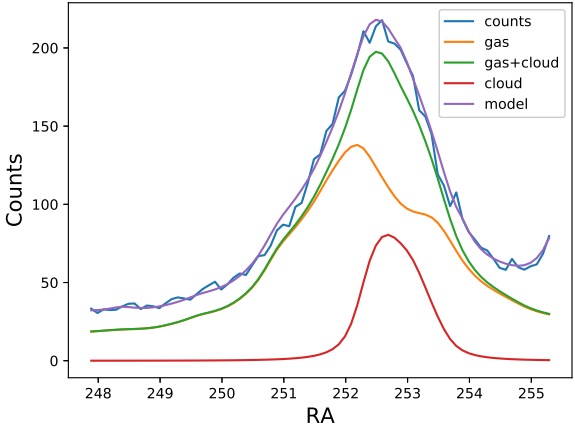}}
\caption{Comparison between the data counts and the model counts of the cloud and of the background  gas, derived {   in the analyzed energy range ($> 800$ MeV)} after the fit, from a slice in DEC correspondent to the extension of the cloud. We show two examples: in (a) cloud 418 that is in the outer Galaxy, in (b) cloud 902 located in the inner Galaxy.} 

\end{figure}
In order to check the results of our likelihood fit, we compared the integrated profile of counts as a function of the right ascension, RA, with the model resulting after the fit, see Fig. 9. The profiles were obtained by integrating over the declination in a slice that corresponded to the cloud extension.  In the plot we show the fitted total model , together with the model of the cloud, the background gas and the sum of the two.  We can see that the total model well represents the observed counts and that the pion decay emission of the gas provides, as expected, the dominant contribution to the observed emission.